# Risk Psychology & Cyber-Attack Tactics

Rubens Kim[1], Stephan Carney[2], Yvonne Fonken[3], Soham Hans[1], Sofia Hirschmann[4], Stacy Marsella[4], Peggy Wu[3], and Nikolos Gurney[1]

[1]Institute for Creative Technologies, University of Southern California, Playa Vista, CA
[2]Marshall School of Business, University of Southern California, Los Angeles, CA
[3]Raytheon Technologies, Farmington, CT
[4]Khoury College of Computer Sciences, Northeastern University, Boston, MA

**ABSTRACT**

We examine whether measured cognitive processes predict cyber-attack behavior. We analyzed data that included psychometric scale responses and labeled attack behaviors from cybersecurity professionals who conducted red-team operations against a simulated enterprise network. We employed multilevel mixed-effects Poisson regression with technique counts nested within participants to test whether cognitive processes predicted technique-specific usage. The scales significantly predicted technique use, but effects varied by technique rather than operating uniformly. Neither expertise level nor experimental treatment condition significantly predicted technique patterns, indicating that cognitive processes may be stronger drivers of technique selection than training or experience. These findings demonstrate that individual cognitive differences shape cyber-attack behavior and support the development of psychology-informed defense strategies.

## INTRODUCTION

Cybersecurity defenses typically focus on technical vulnerabilities—patching software, configuring firewalls, and detecting malicious signatures. However, human attackers remain the ultimate threat, and humans are subject, or vulnerable, to cognitive processes (e.g., biases) that systematically influence decision-making. The IARPA Reimagining Security with Cyberpsychology-Informed Network Defences (ReSCIND) program sought to shift focus from technical countermeasures to exploiting these vulnerabilities to impede attackers. As part of ReSCIND, the Guarding Against Malicious Biased Threats (GAMBiT) project collected multi-modal data from skilled professionals attacking simulated enterprise networks, including responses to psychometric inventories and network activity that was later annotated using a standardized taxonomy of adversarial tactics. Using multilevel modeling, we tested whether differences in cognitive vulnerability predict which tactics individuals employed during the operation.

## PSYCHOLOGICAL VULNERABILITIES AND CYBERSECURITY

The study of cyber attackers' cognitive states, particularly their biases, during an attack is a young field of research. Biases, as studied in judgment and decision making (Tversky & Kahneman, 1974) and occasionally referred to as cognitive vulnerabilities, however, are linked to a multitude of adverse decision outcomes.

Examples are as diverse as under-saving for retirement (Benartzi & Thaler, 2013) and misjudging how charged emotional states will alter your own decision making (Loewenstein & Lerner, 2002). Historically, cybersecurity researchers have focussed on studying the impact of attackers' skills during capture the flag tasks and similar activities—a focus that underweights the cognitive states of cyber attackers (Ferguson-Walter et al., 2019). Early contributions to the research area, such as Ferguson-Walter et al.'s Tularosa Study (Ferguson-Walter et al., 2019), established some pros and cons of experimentally studying attackers' mental states along with exploratory insights into their biases (Gutzwiller et al., 2024). These efforts, along with a related project that pulled together expert opinions on the likely biases experienced by cyber attackers (Johnson et al., 2021) motivated the development of the IARPA ReSCIND program (IARPA, 2025). ReSCIND sought to, "[leverage] well-established cognitive vulnerabilities and human limitations to impede cyber attackers." Accomplishing that meant directly studying attackers' cognitive states, including through the administration of psychological inventories designed to measure cognitive vulnerabilities.

The GAMBiT project conducted experiments in which skilled attackers undertook two eight-hour days of attacking a simulated enterprise network (Beltz et al. 2025). The researchers captured multi-modal data, ranging from self-reports (e.g., participants' hacking skills, demographics, and responses to inventories) to alert data (e.g., Suricata (White et al., 2013)). Full experimental details are available in the Beltz et al. paper, but participants started from a Kali Linux virtual machine (Denis et al., 2016) and pursued objectives laid out in a dossier. Critically, before beginning their attacks, all participants completed four scales designed to measure different aspects of their cognitive tendencies: the Cognitive Reflection Test (CRT, (Frederick, 2005)), the General Risk Propensity Scale (GRiPS, (Zhang et al., 2019)), and two subscales from the Adult Decision Making Competence scale that measure resistance to positive and negative framing (ADMC-RC1 and ADMC-RC2, respectively, (Bruine de Bruin et al., 2007)). Interestingly, the attackers' responses to these scales and their measured attack skills were statistically linked (Oh et al., 2025).

The GAMBiT project also developed an approach for annotating the Suricata logs (Hans, Hirschmann, et al., 2025). This pipeline, which has a large language model (LLM) at its core, was initially developed to understand the notes kept by the attackers (Hans, Gurney, et al., 2025). The annotations are based on the MITRE ATT&CK framework, which is "a globally-accessible knowledge base of adversary tactics and techniques based on real-world observations" used to model threats and provides a clear attack taxonomy (Strom et al. 2020). The effort sought to identify when the attackers 1) noted that they used MITRE ATT&CK persistence techniques and 2) correlate those annotations with the different psychometric scales. The researchers found a statistically significant negative relationship between GRiPS scores and persistence usage, suggesting that attackers with a lower risk propensity relied more on risk-mitigating techniques. We use the Suricata log annotations in our modeling efforts.

**HYPOTHESIS**

We hypothesized that the cyber attackers' cognitive processes, as measured by the scales, would predict the MITRE ATT&CK techniques that attackers use. In other words, the count of each MITRE ATT&CK technique annotated in the individual Suricata logs serve as the dependent variable and the cyber attackers' responses to the various scales (along with experimental controls and demographics) as the independent variables in the modeled relationship.

**METHODS**

We analyzed data from two experiments, GAMBiT Experiments 2 (HSR2, "control," N=17) and 3 (HSR3, "triggers," N=16), conducted under the IARPA ReSCIND program. Data were from cybersecurity professionals who completed datasets a red-team cyber operations in a simulated enterprise network (SimSpace Cyber Force Platform). We excluded data from 31 additional participants due to incomplete data and HSR1 because of changes in the experimental design.

Our data structure consisted of technique counts nested within participants, which are the main dependent variables. The classification of participant's network activity into 11 MITRE ATT&CK technique categories resulted in 363 observations (33 participants × 11 techniques). The 11 MITRE ATT&CK technique categories, derived from Suricata IDS alerts using an LLM annotation pipeline, are: Reconnaissance, Execution, Initial Access, Lateral Movement, Credential Access, Privilege Escalation, Collection, Discovery, Command and Control, Exfiltration, Persistence. Because multiple technique counts from the same individual are not independent, we used multilevel mixed-effects modeling to account for within-participant correlation while testing whether psychometric measures predict technique-specific use patterns.

All 33 participants completed the four psychometric scales, our main independent variables, as part of the experiment prior to engaging in the hackathon. We also rely on other covariates to control for experimental design features, individual differences, and demographic anomalies: Treatment condition (triggers vs. control), Division (Expert vs. Open), age, gender, country (USA vs. other), native English speaker.

We selected Reconnaissance as the reference technique because it had the highest observed frequency (N = 1399, 71.2 % of observations), thus providing the most stable baseline for contrast estimates. Theoretically, reconnaissance represents ongoing information-gathering throughout penetration testing, whereas other techniques (e.g., Initial Access, Lateral Movement) are more discrete events. All coefficients therefore represent effects relative to reconnaissance use.

At Level 1, we modeled technique counts using Poisson regression, appropriate for non-negative count data. The model included main effects for technique categories (with Reconnaissance as the reference due to its high frequency), psychometric × technique interactions to test whether cognitive processes predict techniques differentially, and participant-level covariates (treatment condition, expertise level, demographics). At Level 2, we included participant-level predictors: psychometric scores (ADMC RC1, ADMC RC2, CRT, GRiPS), treatment condition (Triggers vs. Control), expertise level (Expert vs. Open Division), and demographic controls (age, gender, country, native English speaker). All continuous predictors were standardized (M=0, SD=1) prior to analysis to improve numerical stability and facilitate interpretation. We included random intercepts for participants to account for individual differences in overall technique use, allowing each participant to have their own baseline level of activity. This accounts for stable individual differences (e.g., general activity level, overall technical proficiency) while testing technique-specific psychometric effects. Sample size constraints prevented use of random slopes.

The psychometric × technique interactions are the critical feature of our analysis. These interactions test whether specific cognitive processes predict technique use differentially, for example, whether higher reflective thinking (CRT score) predicts more use of some techniques but less use of others.

We fit the models using the lme4 package (version 1.1-37 (Bates et al., 2025)) in R (version 4.5.0) with maximum likelihood estimation and the BOBYQA optimizer (Powell, 2009). Likelihood ratio tests (Type II Wald chi-square) assessed the significance of fixed effects. The model converged with a warning about large eigenvalue ratios, reflecting correlation between the ADMC scales (r=.74) and model complexity. The dispersion ratio (3.64) indicated moderate overdispersion, suggesting unexplained variance beyond that expected under the Poisson distribution. This inflates standard errors (making individual coefficient tests more conservative) but does not invalidate our inference given the strong effects observed (all $p < .002$). +

We report both omnibus likelihood ratio tests for each psychometric × technique interaction and individual coefficient tests for specific technique contrasts. The omnibus tests (one per psychometric measure) control family-wise error rate at α=.05. Individual coefficients are reported for interpretation but should be considered exploratory given the large number of contrasts. Model coefficients represent log rate ratios. For main effects, exp(β) gives the rate ratio for that technique relative to reconnaissance. For interaction terms, exp(β) gives the multiplicative change in the rate ratio associated with a one SD increase in the measure. For example, a coefficient of 0.82 for CRT × Lateral Movement indicates that a one SD increase in reflective thinking is associated with a 2.3-fold increase (exp(0.82)=2.27) in lateral movement relative to reconnaissance.

**RESULTS**

Psychometric measures did not differ significantly between Expert and Open Division groups (all $p > .20$). Across all participants, 1,964 technique uses were identified from network logs, classified into 11 MITRE ATT&CK categories. Reconnaissance dominated (N=1,399, 71%), with all but one participant's Suricata logs showing evidence of this technique. Other frequently used techniques included Lateral Movement (N=206, 13 participants), Execution (N=175, 24 participants), and Initial Access (N=53, 19 participants). Three techniques were rarely observed: Command and Control (3 participants), Exfiltration (3 participants), and Persistence (1 participant).

The intraclass correlation coefficient (ICC) was 0.255, indicating that 25.5% of variance in technique counts was attributable to stable between-person differences, with the remaining 74.5% due to within-person variation across techniques. This substantial between-person variance justified the multilevel modeling approach and indicated meaningful individual differences in overall technique use beyond technique-specific patterns. The model fit the data well (AIC=1797.9, BIC=2039.3). The random intercept standard deviation was 1.06, reflecting considerable heterogeneity in participants' baseline activity levels. Moderate overdispersion was detected (dispersion ratio=3.64), suggesting additional unexplained variance; however, this does not invalidate inference given the extremely strong effects observed (all $p < .002$).

Likelihood ratio tests (Table 1) revealed that all four psychometric measures significantly predicted technique use, but effects varied by technique (see significant psychometric × technique interactions in Table 1). We interpret this as indicating that cognitive processes predict technique preferences differentially rather than uniformly.

**Table 1. Likelihood Ratio Tests**

| Effect | Chi-square | df | p-value |
|---|---|---|---|
| technique | 2137.09 | 10 | < .001 |
| treatment | 1.73 | 1 | 0.189 |
| Division | 0.30 | 1 | 0.584 |
| age_z | 0.16 | 1 | 0.692 |
| gender | 0.78 | 1 | 0.377 |
| country_usa | 1.27 | 1 | 0.260 |
| english_native | 1.65 | 1 | 0.200 |
| technique:ADMC_RC1_z | 122.76 | 11 | < .001 |
| technique:ADMC_RC2_z | 48.42 | 11 | < .001 |
| technique:CRT_z | 198.92 | 11 | < .001 |
| technique:GRiPS_z | 30.72 | 11 | 0.001 |

In contrast, treatment condition (Triggers vs. Control), expertise level (Expert vs. Open Division), and demographic variables did not significantly predict technique use patterns. This indicates that individual cognition, rather than experimental manipulation or experience level, may have been the primary drivers of technique selection in this sample.

To understand these interactions, we examined individual technique-specific effects (See Table 2 for all significant effects with coefficients and confidence intervals. Figures 1 and 2 graphically display the results.). Twelve psychometric × technique combinations showed significant effects at $α=.05$. Higher scores on the resistance to positive framing scale (ADMC RC1) were associated with increased Reconnaissance use, indicating that such individuals engaged in more information-gathering. Conversely, it was negatively associated with Command and Control technique use, though this latter finding should be interpreted cautiously given the rarity of this technique (N=7 total uses).

Higher scores on the resistence to negative framing scale (ADMC RC2) were associated with reduced Reconnaissance use, suggesting that individuals with greater such resistance engaged in less broad information-gathering. It also negatively predicted Lateral Movement, indicating less network traversal among high resistance to negative framing individuals.

Reflective thinking (CRT) showed the most extensive effects, significantly predicting six techniques. Higher CRT was associated with reduced Reconnaissance, Collection, Credential Access, Initial Access, and Privilege Escalation. Notably, CRT showed the opposite effect for Lateral Movement, where more reflective thinkers used this technique significantly more. This suggests that individuals high in reflective thinking employed a qualitatively different tactical approach: less broad reconnaissance and vulnerability scanning, but more strategic network traversal once initial access was achieved.

Risk seeking propensity (GRiPS) positively predicted both Collection and Initial Access techniques. This suggests that individuals who self-reported a higher propensity to accept or seek risks were more likely to engage in data gathering operations and gain system access.

**Table 2: Significant Psychometric × Technique Effects**

| Predictor | Technique | Coefficient | SE | z | p-value |
|---|---|---|---|---|---|
| ADMC_RC1 | Reconnaissance | 1.040 | 0.337 | 3.09 | 0.002 |
| ADMC_RC1 | Command and Control | -2.136 | 0.968 | -2.21 | 0.027 |
| ADMC_RC2 | Reconnaissance | -0.798 | 0.302 | -2.64 | 0.008 |
| ADMC_RC2 | Lateral Movement | -0.760 | 0.325 | -2.34 | 0.019 |
| CRT | Reconnaissance | -0.605 | 0.208 | -2.91 | 0.004 |
| CRT | Collection | -0.685 | 0.315 | -2.17 | 0.030 |
| CRT | Credential Access | -0.861 | 0.240 | -3.59 | < .001 |
| CRT | Initial Access | -0.548 | 0.244 | -2.25 | 0.025 |
| CRT | Lateral Movement | 0.824 | 0.238 | 3.47 | < .001 |
| CRT | Privilege Escalation | -0.646 | 0.298 | -2.17 | 0.030 |
| GRiPS | Collection | 1.117 | 0.441 | 2.54 | 0.011 |
| GRiPS | Initial Access | 1.095 | 0.335 | 3.26 | 0.001 |

**Note: Coefficients for 1 SD increase in predictor.**

In summary, the multilevel analysis revealed robust technique-specific effects of the measured cognitive processes on hacking behavior. Reflective thinking (CRT) showed the most pervasive influence, affecting six different techniques with a consistent pattern: negative effects on reconnaissance and vulnerability exploitation techniques, but a strong positive effect on strategic network traversal. Risk propensity consistently facilitated information gathering and access operations. These differential effects may explain why previous analyses found null results: psychometric effects do not operate uniformly across all techniques but rather shape specific tactical choices within the broader attack process.

We translated selected coefficients into rate ratios to aid interpretation. For example, the CRT × Lateral Movement effect ($\beta=0.82$) indicates that a one SD increase in reflective thinking was associated with a 2.3-fold increase in Lateral Movement use relative to Reconnaissance ($\exp(0.82)=2.27$). Similarly, the strong negative effect of CRT on Credential Access ($\beta=-0.86$) suggests that high reflective thinkers use this technique at only 42% the rate of low reflective thinkers (exp(- 0.86)=0.42), relative to Reconnaissance. These rate differences indicate that cognitive processes can meaningfully influence tactical decisions.

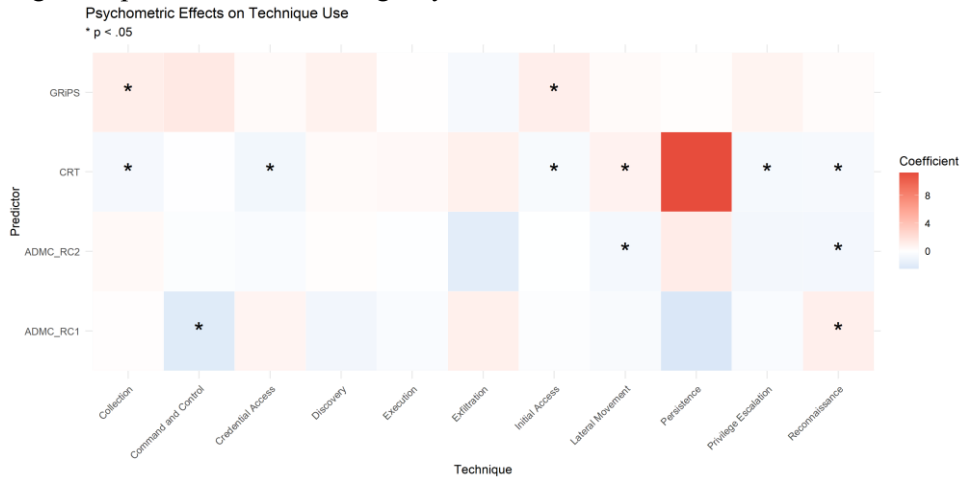

**Figure 1:** A heatmap of all psychometric × technique coefficients. The concentration of negative CRT effects and positive GRiPS effects on specific techniques is visually apparent, as is the relative scarcity of significant effects for rare techniques.

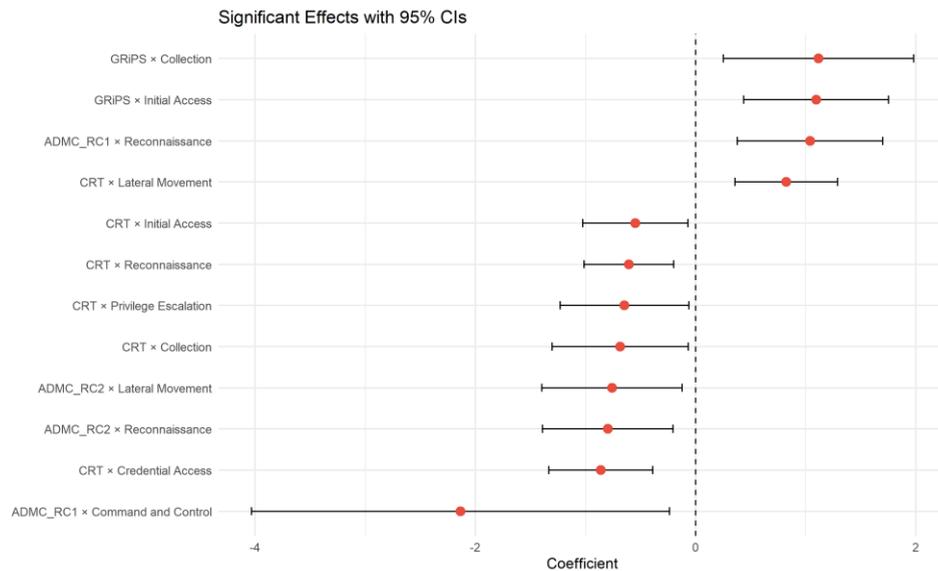

**Figure 2:** The 12 significant effects with 95% confidence intervals. Effect sizes varied considerably, with the largest effects observed for CRT × Lateral Movement (β=0.82) and ADMC RC1 × Command and Control (β=-2.14). The predominantly negative CRT effects and positive GRiPS effects represent the most consistent patterns across techniques.

Despite the significant psychometric effects, the random intercept variance (SD=1.06) indicated substantial individual differences in technique use beyond those explained by measured cognitive processes. Some individuals showed consistently high technique counts across all categories, while others were more selective. This residual heterogeneity suggests that factors beyond the psychometric measures examined here (e.g., strategic preferences, tool familiarity, task interpretation) also influence technique selection.

Notably, neither expertise level (Division) nor experimental treatment (Triggers) significantly predicted technique use patterns. The absence of expertise effects suggests that once individuals reach semi-expert or expert level, technique preferences are possibly more determined by cognitive processes than by training or experience. The null treatment effect indicates that the Triggers manipulation did not substantially alter tactical approaches, though this should be interpreted cautiously given the small sample size.

## DISCUSSION

Our modeling provides evidence that cognitive processes shape cyber attackers' decision-making. All four measures—resistance to framing effects (ADMC RC1 and RC2), reflective thinking (CRT), and risk propensity (GRiPS)—significantly predicted technique-specific usage patterns, supporting our central hypothesis.

The coherent pattern observed for reflective thinking (CRT) suggests that more deliberative attackers adopt a different, more strategic approach. Higher CRT scores predicted less reconnaissance and exploitation but more lateral movement, a pattern consistent with goal-directed behavior over exhaustive enumeration, as described in dual-process theories of cognition (Kahneman, 2011). Similarly, the finding that risk propensity (GRIPS) predicted greater use of collection and initial access techniques aligns with the idea that individuals more comfortable with uncertainty will actively pursue high-value, detectable actions. Conversely, the divergent effects of the two framing resistance subscales (ADMC RC1 and RC2) highlight that different aspects of cognitive control relate to technique preferences in complex, sometimes opposing ways. These nuanced patterns underscore the value of examining multiple cognitive dimensions simultaneously.

Importantly, neither expertise level nor experimental treatment significantly predicted technique usage, suggesting that among skilled practitioners, ingrained cognitive processes may exert a stronger influence on tactical choices than training or experience. This aligns with research showing that expertise does not necessarily reduce vulnerability to cognitive biases (Toet et al., 2016).

Our findings should be considered in light limitations. The overdispersion in our model suggests that unmeasured factors, such as strategic preferences or tool familiarity, also influence technique selection. Our power to detect effects for rarely observed techniques was also limited. Furthermore, while the psychometric scales we employed captured meaningful variance, future work should focus on developing and validating domain-specific cognitive assessments to strengthen the operational relevance of these findings (Flake & Fried, 2020).

## CONCLUSION

This study demonstrates that individual cognitive processes may predict cyber attackers' technique preferences during operations. By employing multilevel modeling with technique-specific interactions, we revealed patterns obscured in previous aggregate analyses. These findings contribute to an emerging cyberpsychology-informed defense paradigm that leverages attackers' cognitive patterns rather than relying solely on technical countermeasures.

Practically, cyber cognitive profiling can be used to predict likely attack paths for resources allocation. For example, attackers high in reflective thinking may warrant monitoring for strategic lateral movement rather than broad scanning activity. Additionally, this work complements concurrent efforts in real-time bias detection (including the detection of attacker ambiguity aversion (Carney et al., 2025), as well as emerging work on a bias-aware Theory of Mind model that improve predictions of boundedly rational adversaries. The convergence of psychological measurement, behavioral modeling, and defensive automation represents a promising frontier for network security.

## ACKNOWLEDGMENT

This research is based upon work supported in part by the Office of the Director of National Intelligence (ODNI), Intelligence Advanced Research Projects Activity (IARPA) under Reimagining Security with Cyberpsychology Informed Network Defenses (ReSCIND) program contract N66001-24-C-4504. The views and conclusions contained herein are those of the authors and should not be interpreted as necessarily representing the official policies, either expressed or implied, of ODNI, IARPA, or the U.S. Government. The U.S. Government is authorized to reproduce and distribute reprints for governmental purposes notwithstanding any copyright annotation therein.